\documentstyle[12pt]{article}

\catcode`\@=11

\def\@normalsize{\@setsize\normalsize{15pt}\xiipt\@xiipt
\abovedisplayskip 14pt plus3pt minus3pt%
\belowdisplayskip \abovedisplayskip
\abovedisplayshortskip  \z@ plus3pt%
\belowdisplayshortskip  7pt plus3.5pt minus0pt}

\def\small{\@setsize\small{13.6pt}\xipt\@xipt
\abovedisplayskip 13pt plus3pt minus3pt%
\belowdisplayskip \abovedisplayskip
\abovedisplayshortskip  \z@ plus3pt%
\belowdisplayshortskip  7pt plus3.5pt minus0pt
\def\@listi{\parsep 4.5pt plus 2pt minus 1pt
            \itemsep \parsep
            \topsep 9pt plus 3pt minus 3pt}}

\def\underline#1{\relax\ifmmode\@@underline#1\else
        $\@@underline{\hbox{#1}}$\relax\fi}
\@twosidetrue

\relax

\catcode`@=12

\catcode`\@=11

\def\ps@headings{\def\@oddfoot{}\def\@evenfoot{}
\def\@oddhead{\hbox{}\hfill
        \makebox[.5\textwidth]{\raggedright\ignorespaces --\thepage{}--
\hfill {}}}}

\def\@evenhead{\@oddhead}
\ps@headings

\catcode`\@=12

\relax

\def\figcap{\section*{Figure Captions\markboth
        {FIGURECAPTIONS}{FIGURECAPTIONS}}\list
        {Fig. \arabic{enumi}:\hfill}{\settowidth\labelwidth{Fig. 999:}
        \leftmargin\labelwidth
        \advance\leftmargin\labelsep\usecounter{enumi}}}
 \relax
\def\tablecap{\section*{Table Captions\markboth
        {TABLECAPTIONS}{TABLECAPTIONS}}\list
        {Table \arabic{enumi}:\hfill}{\settowidth\labelwidth{Table 999:}
        \leftmargin\labelwidth
        \advance\leftmargin\labelsep\usecounter{enumi}}}
 \relax
\def\reflist{\section*{References\markboth
        {REFLIST}{REFLIST}}\list
        {[\arabic{enumi}]\hfill}{\settowidth\labelwidth{[999]}
        \leftmargin\labelwidth
        \advance\leftmargin\labelsep\usecounter{enumi}}}
 \relax

\catcode`\@=11

\def\@evenhead{\@oddhead}

\ps@headings

\relax

\newskip\humongous \humongous=0pt plus 1000pt minus 1000pt

\newif\ifdtup

\def\Tr{\mathop{\rm Tr}}

\def\beq{\begin{equation}}
\def\eeq{\end{equation}}

\def\beqn{\begin{eqnarray}}
\def\eeqn{\end{eqnarray}}
\relax

\def\G2{{\; \rm GeV/}c^2}
\def\G{\; \rm GeV}

\def\dotx{\dotx{\dot\overline{x}}}

\relax

\input {epsf}
\def\p{\partial}
\def\d{{\rm d}}

\def\Lm{\Lambda}
\def\hA{\hat A}

\begin{document}
\begin{titlepage}
\begin{flushright}
       {\normalsize    OU-HET 218  \\  YITP/U-95-29\\
                August,~1995  \\}
\end{flushright}
\vfill
\begin{center}
  {\large \bf  Macroscopic Three-Loop Amplitudes  \\
            and the Fusion Rules \\
            from the Two-Matrix Model }
\footnote{This work is supported in part by
  Grant-in-Aid for  Scientific Research
(07640403)
and by the Grand-in-Aid for Scientific Research Fund
(2690,$~$5108)
from
 the Ministry of Education, Science and Culture, Japan.}
\vfill
           {\bf M. Anazawa}$~^{\dag}$ \footnote{JSPS fellow}
,~~ {\bf A. Ishikawa}$~^{\ddag}$ $~^{2}$
 \\
                          and \\
         {\bf H.~Itoyama}$~^{\dag}$  \\
\vfill
       $~^{\dag}$ Department of Physics,\\
        Faculty of Science, Osaka University,\\
        Toyonaka, Osaka, 560 Japan\\
                       and \\
       $~^{\ddag}$ Uji Research Center, Yukawa Institute for
 Theoretical Physics, \\
        Kyoto University, Uji, Kyoto, 611 Japan \\
\end{center}
\vfill
\begin{abstract}
 From  the computation of  three-point singlet correlators in the two-matrix
 model,  we obtain an explicit expression for the  macroscopic three-loop
 amplitudes having boundary lengths $\ell_{i}$ $(i = 1\sim 3)$
 in the case of the unitary series $(p,q)= (m+1,m)$
  coupled to two-dimensional gravity.
 The sum appearing in this expression
 is found to conform to the structure of the CFT fusion rules
 while the summand  factorizes through a  product of three  modified Bessel
 functions.
   We briefly discuss a possible generalization of these features to
 macroscopic $n$-loop amplitudes.

\end{abstract}
\vfill
\end{titlepage}
The macroscopic $n$-loop amplitudes describe fundamental processes of
$c \leq 1$ noncritical strings.
 Their expression
is well-known
in the case of pure two-dimensional gravity
realized as the critical point of the one-matrix model
\cite{MSS}: it
 takes a suggestive form that the product of factors appears as
 external legs and the rest  are expressible  in terms of
  the sum of the total lengths.  These seem to summarize the lore
 of the noncritical string theory in less than one dimension.
  No comparable expression is known for the case of the unitary
minimal
models
 coupled to gravity
\cite{MSS}, \cite{MS}.
In the previous letter \cite{AII}, however,
 we have obtained
  the two-loop (annulus) amplitude having the boundary lengths
  $\ell_{1}$ and $\ell_{2}$  from
the planar solution of
the two-matrix model for the case of the unitary series;
$(p,q) = (m+1,m)$
\cite{DKK} (See \cite{MSS}, \cite{BDSS},
\cite{AIMZ} for the case of
the one-matrix model).
  Let us first recall this amplitude:
\footnote{
This amplitude is for the case in which the spins on the
two-boundaries are oriented in the same direction.
The case of the opposite direction is obtained by replacing
$\ell_1 + \ell_2$ by $\ell_1 + (-)^m \ell_2$ and by inserting
$(-)^k$ in the summand.}$^,$
\footnote{We have discussed some correspondence
of this amplitude
with the one from the continuum
approach (see also \cite{NKYM}).}
\beqn
\label{result1}
w(\ell_1, \ell_2)_{c} = \frac{2}{m a^2 {\pi}^2}
                    \frac{M \ell_1 \ell_2}{\ell_1 + \ell_2}
                    \sum^{m-1}_{k=1}
                    \left(\sin \frac{k}{m}\pi \right)^2
                    K_{\frac{k}{m}}(M \ell_1)
                    K_{1 - \frac{k}{m}}(M \ell_2)
\quad ,
\eeqn
where $M = t^{1/2}$ and $t$ is the renormalized cosmological
constant (The definition of $a$
and related objects will be given later in eq.~(\ref{3loop-o}),
(\ref{3loop-p}).)
\footnote{A similar but distinct formula is seen in \cite{Kos}.}.
The product of the modified Bessel functions
multiplied by the length
$M\ell_{i} K_{\nu}(M\ell_{i})$
  is naturally interpreted as external leg factors
(wave functions).
This, together with the absence of a propagator, will
constitute integral parts of
Feynman-like rules
 of noncritical $c \leq 1$ string
\footnote{Note that the disk-amplitude is also proportinal to
$(\sin \frac{\pi}{m}) \frac{M^{1+1/m}}{\ell} K_{1+1/m} (M \ell)$.}.
 It is also tempting to regard the integer $k$ summed as a label for
   diagonal primary states of underlying CFT, which is now gravitationally
 dressed.

In this letter, we will corroborate these features further: we
 present an explicit expression for  macroscopic three-loop amplitudes
  in the case of the unitary minimal models coupled to two-dimensional
gravity.
  We obtain this from the computation of correlators consisting
  of the product
 of three resolvents  in the two-matrix model at the $m$-th critical point.
 We first state our expression for the three-loop
 $w(\ell_1,\ell_2,\ell_3)_c$  which is manifestly symmetric under
  the interchange $\ell_{i} \leftrightarrow \ell_{j}$:
\beqn
 w(\ell_1,\ell_2,\ell_3)_c&=& -\frac 1{m(m+1) a^3 \pi^3}
\left(\frac{a M}{2}\right)^{-2-1/m} \nonumber
\\
&\times&
 \sum_{ {\cal D}_{3}(k_{1}-1, k_{2}-1, k_{3}-1)}
 \textstyle{
\sin\frac{k_{1}}m\pi\;\sin\frac{k_{2}}m\pi\;\sin\frac{k_{3}}m\pi}
 \nonumber \\ &\times&
 M^3 \ell_1\ell_2\ell_3\; K_{1-\frac{k_{1}}{m}}(M \ell_1)\;
 K_{1-\frac{k_{2}}{m}}(M \ell_2)\;
 K_{1-\frac{k_{3}}{m}}(M \ell_3)
 \quad .
\label{3loop-final}
\eeqn
  Here  we have denoted by ${\cal D}_{3}$
\beqn
{\cal D}_3 (a_1,a_2,a_3) = \Bigl\{ (a_1,a_2,a_3) \mid
                     \sum^3_{i=1}a_i &\leq& 2(m-2),~~
                     \sum^3_{i=1}a_i = {\rm even},
\nonumber \\
            0 &\leq& \left(\sum^3_{i (\not= j)} a_i \right) - a_j,~~
            j = 1 \sim 3
             \Bigr\}
\eeqn
As one may have expected, the product of three factors
 $(\sin\frac{k_i}{m}\pi)  M\ell_{i}K_{\nu}(M \ell_{i})$ have appeared.

  It is interesting that the selection rules obtained from
  ${\cal D}_{3}$  for the set of integers coincide with the fusion rules
  of the underlying conformal field theory of the unitary minimal series
  $(m+1,m)$ \cite{BPZ}.
In fact, the fusion rules for diagonal primary fields  read as
\beqn
 \langle \phi_{i i}~ \phi_{j j}~ \phi_{k k}\rangle
\not= 0 \quad,
\eeqn
if and only if
$i+j \geq k+1$ and two other permutations and $~i+j+k
{}~(=$ odd) $
\leq 2m-1$ hold.
This set of rules  is nothing but ${\cal D}_{3}(i-1, j-1, k-1)$.

This  coincidence can be understood a little better
  by  studying the  small length behavior of
 eq.~(\ref{3loop-final}).  By letting  $M \ell_{i} << 1$
  independently, we  find that eq.~(\ref{3loop-final}) becomes proportional
to
\beqn
\left(\frac{a M}{2}\right)^{-2-1/m}
 \sum_{ {\cal D}_{3}(k_{1}-1, k_{2}-1, k_{3}-1)}
 M^{\frac{\sum k_{i}}{m}} \ell_{1}^{k_{1}/m} \ell_{2}^{k_{2}/m}
 \ell_{3}^{k_{3}/m} 2^{3-\frac{\sum k_{i}}{m} } \;\;\;,
\label{smallell}
\eeqn
 where we have used
the formula
\beqn
K_{\nu}(x) = \frac{\pi}{2 \sin(\nu \pi)}
\left[
\frac{1}{x^{\nu}} \left\{ \frac{2^{\nu}}{\Gamma(1-\nu)} + O(x^2)
                  \right\}
+ x^{\nu} \left\{ - \frac{1}{2^{\nu} \Gamma(1+\nu)} + O(x^2)
          \right\}
\right]~.
\eeqn
 The $M
$-dependence
  of the individual terms in eq.~(\ref{smallell})
 is $ M^{-2 + \frac{(\sum k_{i})-1}{m}}$.
 This  is compared with the computation done in ref. \cite{DifK}
(See ref. \cite{AGBG} from the point of view of the continuum approach.)
from
 the generalized Kdv  flows
  for  the correlation functions of the dressed primaries:
\beqn
 \langle
 \phi_{\ell}~ \phi_{p}~ \phi_{r}
\rangle
 = \ell p r ~(\frac {u}{2})^{(\ell + p + r - 3)/2}
 u^{\prime} \;\;.
\eeqn
  Under $u~
(= t^{1/m})
$, the $t$-dependence of this expression completely
 agrees with that of eq.~(\ref{smallell}).  The selection rules  for this
 expression found in ref. \cite{DifK} is again summarized as
  $ {\cal D}_{3}$.

Our formula   tells how
  the higher order operators (gravitational descendants) in addition to the
dressed primaries are constrained to
  obey   the selection rules of CFT.
Let us remark here that
 the $c=1$ limit can be studied from our formula
\cite{MS}.
We have checked that, by letting one of the three lengths shrink
($M \ell_i<<1$),
the three-loop amplitude
reduces to the derivative of the two-loop amplitude
with respect to the cosmological constant $t$
up to a normalization factor.


  The remainder of this letter is devoted to the derivation of
 eq.~(\ref{3loop-final})
  from the two-matrix model at the $(m+1, m)$ critical point.
   We will also briefly discuss  what in eq.~(\ref{3loop-final}) may
 be generalized  to the expression for macroscopic $n$-loop amplitudes.

To begin with
let us consider the connected three-point  singlet correlator
  consisting of   arbitrary analytic
 functions $f(\hA)$, $g(\hA)$, and $h(\hA)$,
where the matrix $\hA$ is a dynamical variable.
In the bases of orthogonal polynomials,
one can derive
\beqn
 \langle \Tr f(\hA)\;\Tr g(\hA)\; \Tr h(\hA) \rangle _c
 &=& \sum_{i=0}^{N-1} \sum_{k=N}^{\infty} \sum_{l=N}^{\infty}
 \;[f(A)]_{ik} \;[g(A)]_{kl} \;[h(A)]_{li} \nonumber \\
 &-& \sum_{i=0}^{N-1} \sum_{k=N}^{\infty} \sum_{l=0}^{N-1}
 [f(A)]_{ik} \;[h(A)]_{kl} \;[g(A)]_{li}
 \quad ,
\label{3loop-a}
\eeqn
where
$
 [f(A)]_{ik} \equiv \langle i| f(\lambda) |k \rangle
$.
It is convenient to introduce the following `classical' function in the case
only the planar limit is of interest,
\beq
 A(z,s,\Lambda) \equiv \sum_k z^k\; A(i)_k \quad ,\quad
 s \equiv \Lambda i/N
 \quad ,
\label{3loop-c}
\eeq
where
$
 A(i)_{k} \equiv A_{i-k,i}
$.
The `classical' function depends on the
 bare cosmological constant $\Lambda$ only
through $s$ when we take $N\to\infty$ limit.
It is, therefore, legitimate
to introduce
$
\displaystyle
 A(z,s) \equiv \lim _{N\to \infty }A(z,s,\Lambda)
:$
\beq
 A(z,s,\Lambda) = A(z,s) + O(1/N)
\quad .
\label{3loop-f}
\eeq
In terms of the `classical' function, we have
\beq
 [f(A)](N)_k =\frac 1{2\pi i} \oint \frac{\d z}{z^{k+1}} f\left(
 A(z,s=\Lambda)\right) + O(1/N)
\label{classical}
\quad ,
\label{3loop-g}
\eeq
where
$
 [f(A)](i)_k \equiv [f(A)]_{i-k,i}
$.
In the right-hand sind of eq.~(\ref{3loop-a}),
the leading terms in $1/N$ of the first term and
those of the second term
get cancelled. We have to consider
the
next leading
terms. For any integer $\epsilon$, we obtain
\beqn
 [f(A)](N+\epsilon)_k &=& \frac 1{2\pi i} \oint\frac{\d z}{z^{k+1}} f\left(
 A(z,s=\Lambda)\right) \nonumber\\
 &+& \frac{\Lambda\epsilon}N \frac 1{2\pi i} \oint\frac{\d z}{z^{k+1}} \left.
 \frac{\p A(z,s)}{\p s} \right|_{s=\Lambda}
 \frac{\p f\left(A(z,\Lambda)\right)} {\p A} \nonumber \\
 &+&(\mbox{the part independent of $\epsilon$}) +O(1/N^2)
 \quad .
\label{3loop-i}
\eeqn
The part independent of $\epsilon$ comes from the terms $O(1/N)$ in
eq.~(\ref{3loop-g}). The second term is
responsible for the computation in what follows.
Using
eq.~(\ref{3loop-i})
and considering the terms $1/N$ in eq.~(\ref{3loop-a}), we obtain
\beqn
 \lefteqn{
 \langle \Tr f(\hA)\; \Tr g(\hA)\; \Tr h(\hA) \rangle _c
 } \nonumber \\
 &=&
 \frac{\Lambda}N \sum_{\delta_1=0}^{\infty}\sum_{\delta_2=0}^{\infty}
 \sum_{\delta=0}^{\infty } \nonumber \\
 &&
 \left\{ (\delta_2-\delta_1)\;\frac{\p}{\p s}[f(A)](N)_{\delta_1+\delta_2+1}\;
 [g(A)](N)_{\delta-\delta_2}\; [h(A)](N)_{-\delta-\delta_1-1} \right.
 \nonumber \\
 &+&  (\delta+\delta_2+1)\; [f(A)](N)_{\delta_1+\delta_2+1}\; \frac{\p}{\p s}
 [g (A)](N)_{\delta-\delta_2}\; [h(A)](N)_{-\delta-\delta_1-1} \nonumber \\
 &+& \left. (\delta-\delta_1)\; [f(A)](N)_{\delta_1+\delta_2+1}\;
 [g(A)](N)_{\delta-\delta_2}\; \frac{\p}{\p s}[h(A)](N)_{-\delta-\delta_1-1}
 \; \right\} \nonumber \\
 &+& O(1/N^2)
 \quad .
\label{3loop-j}
\eeqn
Using eq.~(\ref{classical})
the three-point function eq.~(\ref{3loop-a}) in the planar
limit can
be expressed in terms of the `classical' function as
\beqn
\lefteqn{
\langle \Tr f(\hA)\; \Tr g(\hA)\; \Tr h(\hA) \rangle _c } \nonumber \\
 &=&
 \frac\Lambda N \frac 1{(2\pi i)^3} \oint_{|z|>|z'|>|z''|} \d z\d z'\d z''
 \nonumber \\ &\mbox{}&
 \left\{ \frac z{(z-z')^2(z-z'')^2}f'\left(A(z)\right)
\frac{\p A(z)}{\p\Lambda}
 g\left(A(z')\right) h\left(A(z'')\right) \right. \nonumber \\
 &+&
 \frac {z'}{(z-z')^2(z'-z'')^2} f\left(A(z)\right)
 g'\left(A(z')\right)\frac{\p A(z')}{\p\Lambda} h\left(A(z'')\right)
 \nonumber \\
 &+& \left.
 \frac {z''}{(z-z'')^2(z'-z'')^2} f\left(A(z)\right)
 g\left(A(z')\right) h'\left(A(z'')\right)\frac{\p A(z'')}{\p\Lambda} \right\}
 \quad ,
\label{3loop-k}
\eeqn
where we set $A(z) \equiv A(z, s = \Lambda)$.
 From the above formula (eq.~(\ref{3loop-k})), the three-point resolvent in
the planar limit is expressed as
\beqn
 \lefteqn{
 \frac N\Lambda \Bigl\langle \Tr\frac 1{p_1-\hA}\; \Tr\frac 1{p_2-\hA}\;
 \Tr\frac 1{p_3-\hA} \Bigr\rangle _c  } \nonumber \\
 &=&
 \frac 1{(2\pi i)^3} \oint_{|z|>|z'|>|z''|} \d z\d z'\d z''
 \nonumber \\ &\mbox{}&
 \left\{ \frac z{(z-z')^2(z-z'')^2} \frac{\p
A(z)}{\p\Lm}\frac{1}{[p_1-A(z)]^2}
 \frac{1}{p_2-A(z')} \frac{1}{p_3-A(z'')} \right. \nonumber \\
 &+&
 \frac {z'}{(z-z')^2(z'-z'')^2} \frac{1}{p_1-A(z)} \frac{\p A(z')}{\p\Lm}
 \frac{1}{[p_2-A(z')]^2}\frac{1}{p_3-A(z'')} \nonumber \\
 &+& \left.
 \frac {z''}{(z-z'')^2(z'-z'')^2} \frac{1}{p_1-A(z)} \frac{1}{p_2-A(z')}
 \frac{\p A(z'')}{\p\Lm}\frac{1}{[p_3-A(z'')]^2} \right\}
 \quad ,
\label{3loop-l}
\eeqn
where the contour of $z$ encircles that of $z'$
and similarly the contour of $z'$ encircles that of $z''$.
We calculate these contour integrals and find
\beqn
 \frac N\Lambda \Bigl\langle \Tr\frac 1{p_1-\hA} &\Tr&\frac 1{p_2-\hA}\;
 \Tr\frac 1{p_3-\hA} \Bigr\rangle _c \nonumber \\
 &=&
 -\frac{1}{a^3} \left\{
 \frac{\p z_1}{\p\Lm} \frac{\p}{\p z_1}
 \left[ \frac{z_1}{(z_1-z_2)^2(z_1-z_3)^2}\frac{\p z_1}{\p \zeta_1}\right]
 \frac{\p z_2}{\p \zeta_2}\frac{\p z_3}{\p \zeta_3} \right. \nonumber \\
 &+&
 \frac{\p z_1}{\p \zeta_1}
 \frac{\p z_2}{\p\Lm} \frac{\p}{\p z_2}
 \left[ \frac{z_2}{(z_1-z_2)^2(z_2-z_3)^2}\frac{\p z_2}{\p \zeta_2}\right]
 \frac{\p z_3}{\p \zeta_3}  \nonumber \\
 &+& \left.
 \frac{\p z_1}{\p \zeta_1}
 \frac{\p z_2}{\p \zeta_2}
 \frac{\p z_3}{\p\Lm} \frac{\p}{\p z_3}
 \left[ \frac{z_3}{(z_1-z_3)^2(z_2-z_3)^2}\frac{\p z_3}{\p \zeta_3}\right]
 \;\right\}
 \quad ,
\label{3loop-m}
\eeqn
where
\beq
 z_i=\exp (2\eta \cosh \theta_i)
\label{3loop-n}
\eeq
\beq
p_i - p_* =
 a \zeta_i=2 \eta^m \cosh m\theta_i=a M \cosh m\theta_i \quad,\quad
 \eta=(a M/2)^{1/m}
\label{3loop-o}
\eeq
\beq
 \Lm-\Lm_*=-(m+1) \eta^{2m}=-(m+1)a^2 \mu =-(m+1)(aM/2)^2
\label{3loop-p}
\eeq
and $p_*, ~\Lambda_*$ denote the critical values of $p, ~\Lambda$
respectively.
The right hand side of eq.~(\ref{3loop-m}) can be written in a compact form as
\beqn
 \frac N\Lambda \Bigl\langle \Tr\frac 1{p_1-\hA} &\Tr&\frac 1{p_2-\hA}\;
 \Tr\frac 1{p_3-\hA} \Bigr\rangle _c \nonumber \\
 &=&\frac 1{2^2 m(m+1)a^3}\left(\frac{aM}2\right)^{-2-1/m}
\frac{\p}{\p\zeta_1}
 \frac{\p}{\p\zeta_2}\frac{\p}{\p\zeta_3} F(\theta_1,\theta_2,\theta_3)
 \quad ,
\label{3loop-q}
\eeqn
where
\beqn
 F(\theta_1,\theta_2,\theta_3)
 &=&
 \frac{1}{(\cosh\theta_1-\cosh\theta_2)(\cosh\theta_1-\cosh\theta_3)}
 \;\frac{\sinh (m-1)\theta_1}{\sinh m\theta_1} \nonumber\\
 &+& \quad (1\leftrightarrow 2) \quad + \quad (1\leftrightarrow 3)
 \quad .
\label{3loop-r}
\eeqn
 Making use of the following formula twice,
\beqn
 \lefteqn{
 \frac{1}{\cosh\alpha-\cosh\beta}
 \left(\;\frac{\sinh (n-k)\alpha}{\sinh n\alpha}
 -\frac{\sinh (n-k)\beta}{\sinh n\beta} \;\right) \mbox{                 }
 } \nonumber\\
 &&=-2\sum_{j=1}^{n-k}\sum_{i=1}^{k}
 \;\frac{\sinh (n-j-i+1)\alpha}{\sinh n\alpha}
 \;\frac{\sinh (n-j-k+i)\beta}{\sinh n\beta}
 \quad ,
\label{3loop-s}
\eeqn
  we find that
eq.~(\ref{3loop-r})
is written as a triple sum where the summand factorizes
into three factors associated with individual loops:
\beqn
 \lefteqn{
 F(\theta_1,\theta_2,\theta_3) } \nonumber \\
 &=&
 4\sum_{k=1}^{m-1}\sum_{j=1}^{m-k}\sum_{i=1}^{k}
 \;\frac{\sinh (m-k)\theta_1}{\sinh m\theta_1}
 \;\frac{\sinh (m-j-i+1)\theta_2}{\sinh m\theta_2}
 \;\frac{\sinh (m-k-j+i)\theta_3}{\sinh m\theta_3} \nonumber\\
 &&
 \quad .
\label{3loop-t}
\eeqn
Here, $W(\zeta_1, \zeta_2 , \zeta_3)_c
= \frac N\Lambda \Bigl\langle \Tr\frac 1{p_1-\hA}\; \Tr \frac 1{p_2-\hA}
 \; \Tr\frac 1{p_3-\hA} \Bigr\rangle _c
 $
and $w(\ell_{1}, \ell_{2}, \ell_{3})_c$ are related
by the Laplace transform
\beqn
\label{Laplace}
W(\zeta_1, \zeta_2, \zeta_3)_c &=& \int^{\infty}_{0} d \ell_{1}
                             \int^{\infty}_{0} d \ell_{2}
                             \int^{\infty}_{0} d \ell_{3}
                             e^{- \zeta_1 \ell_1} e^{- \zeta_2 \ell_2}
                             e^{- \zeta_2 \ell_3}
                             w(\ell_{1}, \ell_{2}, \ell_{3})_c
\;\;\; \\ \nonumber
&\equiv& {\cal L}[w(\ell_{1}, \ell_{2}, \ell_{3})_c] \quad .
\eeqn
  In \cite{AII},
we have found the following formula for the inverse Laplace image
\beqn
\label{inverse}
{\cal L}^{-1}[\frac{\partial}{\partial \zeta}
         \frac{\sinh k \theta}{\sinh m \theta}]
= - \frac{M \ell}{\pi} \sin \frac{k \pi}{m}~  K_{\frac{k}{m}} (M \ell)
\quad .
\eeqn
Note that $K_{\nu}(z)$ is the modified Bessel function.


Inverse Laplace transform of eq.~(\ref{Laplace})
gives the three-loop amplitude in terms of
each boundary length.  We obtain
\beqn
 w(\ell_1,\ell_2,\ell_3)_c&=& -\frac 1{m(m+1) a^3 \pi^3}
\left(\frac{a M}{2}\right)^{-2-1/m} \nonumber
\\
&\times&
 \sum_{k=1}^{m-1}\sum_{j=1}^{m-k}\sum_{i=1}^{k}
 \textstyle{
\sin\frac{m-k}m\pi\;\sin\frac{m-j-i+1}m\pi\;\sin\frac{m-k-j+i}m\pi}
 \nonumber \\ &\times&
 M^3 \ell_1\ell_2\ell_3\; K_{\frac{m-k}{m}}(M \ell_1)\;
 K_{\frac{m-j-i+1}{m}}(M \ell_2)\;
 K_{\frac{m-k-j+i}{m}}(M \ell_3)
 \quad .
\label{3loop-u}
\eeqn

  Let us write  the set of restrictions on the triple sum in
 eq.~(\ref{3loop-u}) as
\beqn
{\cal F}(k,j,i) = \Bigl\{
                         (k,j,i) \mid 1 \leq k \leq m-1,~
      1 \leq j \leq m-k,~ 1 \leq i \leq k
                  \Bigr\} \quad .
\eeqn
 By  elementary algebras, one can show
\beqn
{\cal F}(k,j,i) &=& {\cal D}_3 \left(k_1 -1,~ k_2 -1,~ k_3 -1 \right)
\nonumber \\
&=& {\cal D}_3 \left(k-1,~(j+i-1)-1,~(k+j-i)-1 \right)
\quad .
\eeqn
 This completes the derivation of eq.~(\ref{3loop-final}).

  Finally, let us briefly discuss how the
features of the three-loop amplitude
  we have found here may be generalized to the $n$-loop amplitudes.
  It is for sure that  the summand contains  a product of $n$-external leg
 factors $(\sin\frac{k_{i}}m\pi)
 M \ell_{i} K_{1-\frac{k_{i}}{m}}(M \ell_{i})$
  $(i=1 \sim m)$  but
 it must be supplemented by a symmetric polynomial, which we need to
 determine.
 As for the restrictions on the sum, let us first note
 that ${\cal D}_{3}$, which is fusion rules,
is rephrased as
  the condition that a triangle with edges $(a_{1}, a_{2}, a_{3})$
 be formed whose circumference is even
 and less than or equal to $2(m-2)$.
The $n$-loop amplitudes must respect the fusion rules and
these triangles
will become building blocks  for the case of  $n$-loop amplitudes.
 The $n$-gons made out of $n-2$ triangles of this type
\beqn
{\cal D}_{n}(a_1,....,a_n)  = \Bigl\{ (a_1,....,a_n) \mid
                       \sum^n_{i=1} a_i = {\rm even},
\nonumber \\
            0 \leq \left(\sum^n_{i \not= j} a_i \right) - a_j,~~
            j = 1 \sim n
               \Bigr\}
\eeqn
 are relevant.   Different shapes of $n$-gons for given
  $(a_1,....,a_n)$ and various different divisions
  of each $n$-gon  are important ingredients of the
  completely and crossing symmetric properties of the amplitudes.
  Also  note that the restriction on the sum in the case
 of the two-loop (annulus)  amplitudes is written as
 ${\cal D}_{2}(m-k_{1}, k_{2})$ under the convention of the factors
  $ M^2 \ell_1\ell_2 K_{1-\frac{k_{1}}{m}}(M \ell_1)\;
 K_{1-\frac{k_{2}}{m}}(M \ell_2)$.
   Despite these prominent features,   complete determination
 of the amplitudes still requires  computation from the two-matrix model.



\newpage

\begin{thebibliography}{99}

\bibitem{MSS}
G.~Moore, N.~Seiberg and M.~Staudacher,
{\sl Nucl. Phys.}~{\bf B362}~(1991)~665.

\bibitem{MS}
G.~Moore and N.~Seiberg,
{\sl Int. J. Mod. Phys.}~{\bf A7}~(1992)~2601.

\bibitem{AII}
M.~Anazawa, A.~Ishikawa and H.~Itoyama,
preprint OU-HET 190, YITP/U-94-34,
hep-th/9410015,
to appear in {\sl Phys. Rev.}~{\bf D}.

\bibitem{DKK}
J.M.~Daul, V.A.~Kazakov and I.K.~Kostov,
{\sl Nucl. Phys.}~{\bf B409}~(1993)~311.

\bibitem{BDSS}
T.~Banks, M.~Douglas, N.~Seiberg and S. Shenker,
 {\sl Phys. Lett.}~{\bf B238}~(1990)~279.

\bibitem{AIMZ}
L.~Alvarez-Gaum\'{e}, H.~Itoyama, J.L.~Manes and A.~Zadra,
{\sl Int. J. Mod. Phys.}~{\bf A7}~(1992)~5337.

\bibitem{NKYM}
  R.~Nakayama, {\sl Phys. Lett.}~{\bf B325}~(1993)~347.

\bibitem{Kos}
 I. K.~Kostov, {\sl Nucl. Phys.}~{\bf B376}~(1992)~539.

\bibitem{BPZ}
A.~Belavin, A.M.~Polyakov and A.B.~Zamolodchikov,
{\sl Nucl.~Phys.}~{\bf B241}    (1984)~333.


\bibitem{DifK}
   P.~DiFrancesco and D.~Kutasov,
{\sl Nucl. Phys.}~{\bf B342}~(1990)~589.

\bibitem{AGBG}
  L.~Alvarez-Gaume, J.~Barbon and C.~Gomez,
 {\sl Nucl. Phys.}~{\bf B368}~(1992)~57.

\end{thebibliography}
\end{document}